\documentclass[lettersize,journal]{IEEEtran}
\usepackage{algorithmic}
\usepackage{url}
\urldef{\urlA}\url{https://github.com/tudtude/MICROSERVICE-RESTful-Nodejs} %
\urldef{\urlB}\url{https://github.com/dorsal-lab/tracecompass_nodejs\_support} 
\usepackage{algorithm}
\usepackage{array}
\usepackage[caption=false,font=normalsize,labelfont=sf,textfont=sf]{subfig}
\usepackage{textcomp}
\usepackage{stfloats}
\usepackage{verbatim}
\usepackage{amssymb}
\usepackage[numbers]{natbib}
\usepackage{cite}
\usepackage{amsmath,amssymb,amsfonts}
\usepackage{graphicx}
\usepackage{subfig}
\usepackage{multirow}
\usepackage{listings}
\usepackage{xcolor}
\definecolor{codegreen}{rgb}{0,0.6,0}
\definecolor{codegray}{rgb}{0.5,0.5,0.5}
\definecolor{codepurple}{rgb}{0.58,0,0.82}
\definecolor{backcolour}{rgb}{0.95,0.95,0.92}

\begin{document}
\title{Advanced Strategies for Precise and Transparent Debugging of Performance Issues in In-Memory Data Store-Based Microservices}
\author{Herve M. Kabamba, Matthew Khouzam,  Michel R. Dagenais}


\maketitle

\title{Advanced Strategies for Precise and Transparent Debugging of Performance Issues in In-Memory Data Store-Based Microservices}

\author{
    \IEEEauthorblockN{Herve M. Kabamba\IEEEauthorrefmark{1}, Michel Dagenais\IEEEauthorrefmark{1}, Matthew Khouzam\IEEEauthorrefmark{2}}
}

\begin{abstract}
The rise of microservice architectures has revolutionized application design, fostering adaptability and resilience. These architectures facilitate scaling and encourage collaborative efforts among specialized teams, streamlining deployment and maintenance. Critical to this ecosystem is the demand for low latency, prompting the adoption of cloud-based structures and in-memory data storage. This shift optimizes data access times, supplanting direct disk access and driving the adoption of non-relational databases.

Despite their benefits, microservice architectures present challenges in system performance and debugging, particularly as complexity grows. Performance issues can readily cascade through components, jeopardizing user satisfaction and service quality. Existing monitoring approaches often require code instrumentation, demanding extensive developer involvement. Recent strategies like proxies and service meshes aim to enhance tracing transparency, but introduce added configuration complexities.

Our innovative solution introduces a new framework that transparently integrates heterogeneous microservices, enabling the creation of tailored tools for fine-grained performance debugging, especially for in-memory data store-based microservices. This approach leverages transparent user-level tracing, employing a two-level abstraction analysis model to pinpoint key performance influencers. It harnesses system tracing and advanced analysis to provide visualization tools for identifying intricate performance issues. In a performance-centric landscape, this approach offers a promising solution to ensure peak efficiency and reliability for in-memory data store-based cloud applications.
\end{abstract}


\section{Introduction}
\label{sec:intro}
\IEEEPARstart{T}he advent of microservice architectures has changed the paradigm with which applications were once designed to adopt a much more resilient architecture. These architectures incorporate a simplistic scaling dimension that allows for adequate adjustments to the system, without disrupting its operation. They have encouraged collaboration among different teams responsible for specific components of the system, facilitating deployment, development, and maintenance.

A critical requirement, for the overwhelming majority of applications functioning within this ecosystem, is minimal latency, \citet{cho2013instant}. Presently, the user experience has been substantially enhanced as a result of the swift progressions in technology. The utilization of cloud-based architectures has necessitated the placement of data storage in memory, due to the latency restrictions imposed by applications. This paradigm has significantly decreased data access times, due to the fact that data is now accessible at the memory level, rather than directly from the disk. In addition, the emergence of Big Data and Cloud Computing has prompted numerous organizations to forgo relational databases in favor of non-relational ones \citep{seghier2021performance}.

While these architectures have brought greater flexibility to the management and development of computer systems, they have also raised numerous challenges. Among them, system performance remains a crucial element, while performance debugging increases in complexity, as explained by \citet{Maximilian2020}.

Common approaches to performance issue debugging typically employ application instrumentation methods to collect execution traces. A notable challenge with these approaches is the necessity to modify the application source code to insert tracepoints at specific locations, often resulting in altered system behavior. This method has become widely prevalent in the microservices ecosystem, particularly with the use of distributed tracing.

Despite the advantages offered by these approaches, they have limitations in terms of granularity. Specifically, distributed tracing instrumentation allows for tracepoints placement only on microservice components, neglecting the backends connected to them. The storage component plays a crucial role in microservice architectures, as it facilitates data persistence or access, and can be a significant source of performance degradation.

High-performance architectures typically incorporate in-memory databases for rapid data access, messaging services, and even caching services. Usually, requests to these systems are instrumented at the microservice level, to measure latency. However, when these systems are the root cause of performance issues, distributed tracing renders them as black boxes, making it virtually impossible to trace back to the internal source of the problem.

Debugging such issues requires direct instrumentation of these systems, necessitating significant advancements in source code understanding and trace analysis development. Methods that enable developers to transparently trace microservices and various backends, particularly those providing the storage dimension to the architecture, are crucial. These methods help reduce costs and facilitate the debugging of performance issues in microservices.

\citet{menasce2002qos} and \citet{luk2005pin}, along with other scholars, have put forth methods to address the typically high costs associated with instrumentation efforts.

Similarly, recent approaches have employed proxies that establish an intermediary layer within the traced system \citep{santana2019transparent}. This intermediary layer intercepts system calls generated during communications between subsystems at the lowest level. 

We propose a novel approach to performance analysis in microservic architectures that incorporates the storage dimension. This approach aims to achieve two primary objectives: First, it involves instrumenting the in-memory data-store, Redis in our case, to develop an analysis that precisely debugs internal performance issues. Second, it offers a method for seamlessly integrating various analyses into a unified system, providing a comprehensive view of the whole architecture, and facilitating the identification of bottlenecks. Recognizing these bottlenecks enables the use of granular analyses previously developed to pinpoint these issues.

The integration is exemplified using two distinct analyses. The first, derived from previous work \citep{mbikayi2023toward}, involves the transparent tracing of microservices developed in \texttt{Node.js} \citep{nodejs}. This process requires no additional instrumentation effort from the developer. Instead, the developer merely deploys the instrumented image of the \texttt{Node.js} environment with the application in a container, then applies automated analysis to the obtained trace to visualize microservice interactions.

The second analysis results from the direct instrumentation of \texttt{Redis}, chosen for its popularity and its encompassing functionalities as a cache, database, and messaging service. The same method could similarly be applied to other in-memory data-store systems.

Our approach enables transparent tracing in these two environments, reconstructing interactions through automated trace analyses, and visualizing interactions between different components. In other words, it facilitates a comprehensive and granular performance analysis of the entire system.

Furthermore, this approach can be expanded to other environments, allowing the integration of environments beyond \texttt{Node.js} and \texttt{Redis}.

The following are the primary contributions of this paper:
(1) We introduce an integrated framework for performance debugging in microservice architectures. 
(2) We propose an abstraction model that captures actors likely to influence in-memory data-store performance, and demonstrate how such a model can lead to fine-grained performance debugging. This model can be applied to other systems.
(3) We demonstrate the relevance and effectiveness of our framework through practical and complex use cases.

The remainder of the paper is organized as follows: Section~\ref{sec:related} discusses related work on the subject. Section~\ref{sec:method} describes the proposed framework. Section~\ref{sec:exp} describes the context of our experiments, and present some use cases to support the relevance of our tool. In Section~\ref{sec:eval}, an evaluation of the solution is performed,  while in Section~\ref{sec:conclusion} a conclusion on the work is drawn. 
\section{Related Work}
\label{sec:related}
Developers spend the majority of their time debugging problems in applications as explained in \citep{complete2004steve}.

 \citet{gan2019seer}, porposed Seer, an online debugger that predicts quality of service violations in cloud applications. The study is based on a microservices infrastructure that uses Memcached, which has similarities to \texttt{Redis} as an in-memory database. 

Debugging performance issues with Seer, involves an instrumentation step of microservices. It is also performed in the Memcached data store, mainly at the polling functions and various network interface queues. Seer is capable of predicting quality of service violations based on a model generated using deep learning on upstream traces. Instrumenting the functions that manage packet queuing, at the data store level, allows for identifying bottlenecks when they involve Memcached.

\citet{merino2023microservice} proposed a debugging approach that utilizes C/R (Checkpoint/Restore) techniques with CRIU (Checkpoint/Restore in Userspace). This method aims to replicate faulty environments, mitigating the iterative nature of the debugging process. Their approach extends the checkpoint capabilities of C/R engines based on ptrace, providing support for containerized processes attached to the debugger, all without the need for code instrumentation or custom operating system kernels.

OmniTable, a query model used as a debugging approach was proposed in\citep{quinn2022debugging}. It aims to reduce programming complexity and minimize the performance overhead associated with debugging, while preserving for developers an unrestricted access to the execution state. They demonstrated that this query model simplifies the debugging query formulation process, compared to existing tools. The approach was tested through a series of case studies involving well-known open-source software such as \texttt{Redis}. Their approach produces a model of the system states that can be replayed to understand its behavior. However, an issue arises with the size of the tree structure that needs to be stored in memory. It relies on logging techniques to capture the execution history associated with each OmniTable, and then generates the instrumentation and traces needed for a new replay execution. It has been applied to a specific \texttt{Redis} use case to identify a performance issue. However, this approach is based on a model that requires significant resources.

\citet{whittaker2018debugging}, proposed Wat-provenance to leverage a data provenance approach to debug performance issues. It takes into account the causality dimension. A state machine is used to describe why a particular data is produced as an output.  The approach was applied in a distributed systems context for root causes inference. The distributed system infrastructure included \texttt{Redis} servers in their experiments. However, such an approach can only infer external communication behavior and bottleneck, but cannot be applied to debug the internal \texttt{Redis} components functioning.
\citet{arafa2017qdime}, proposed a dynamic instrumentation approach for performance debugging. Extraction of runtime information was applied to comprehend system behavior. The approach was tested on \texttt{Redis}.

\citet{santana2019transparent} introduced an innovative method for achieving transparent tracing, which harnesses the kernel of the operating system to capture system calls related to communication among microservices. They advocated the use of a proxy that adds a neutral layer to the microservice, enabling the interception of its interactions and the correlation of information to infer causality associated with different requests. Intercepting system calls ensures transparency in application tracing, but it entails developer responsibility for configuring the infrastructure.

For service discovery, \citet{wassermann2011monere} employed dependency structures extracted from documentation using statistical methods. In addressing fault detection, \citet{chen2002pinpoint} used middleware instrumentation to log the specific components handling individual requests. While these approaches achieved a certain level of transparency, they demanded extensive dedication of the developer time for library instrumentation.

Distributed tracers such as Dapper, proposed by \citet{sigelman2010dapper}, and X-trace, introduced by \citet{fonseca2007x}, are capable of tracing the entire request lifecycle, revealing its flow, and assisting in diagnosing issues throughout execution. They rely on mechanisms for injecting a trace context to reconstruct the trace nesting structure. However, these approaches require an initial instrumentation phase, for the application to activate the collection mechanism. 

Tracing strategies for request paths were also addressed by \citet{kitajima2017inferring} using heuristics. \citet{aguilera2003performance} proposed algorithms for diagnosing request causality. Both of these approaches offer a certain degree of transparency but rely on middleware instrumentation.

Unlike most of these approaches, our framework refrains from injecting trace context into the request; instead, it capitalizes on the internal mechanism of host environments, to reconstruct the request path.

\section{Proposed Framework}
\label{sec:method}
\subsection{Integrated Framework Architecture}
\begin{figure*}[ht]
\centering{\includegraphics[width=10cm,height=7cm]{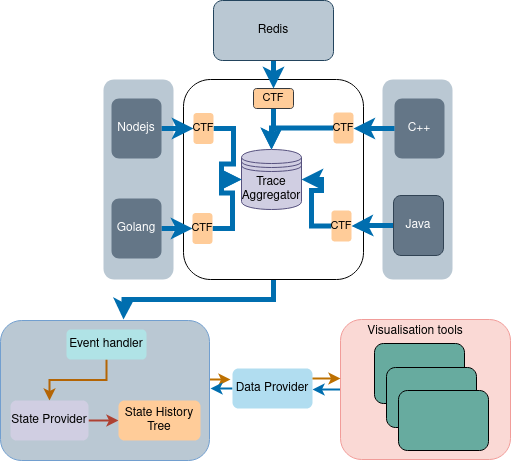}} 
\caption{Performance Analysis Integration Framework architecture}

\label{fig:architecturef}
\end{figure*}

Figure~\ref{fig:architecturef} shows the overall architecture of the integration framework that our approach proposes. The various microservice environments, in their heterogeneity, are considered building blocks for constructing a performance analysis framework that takes into account the entirety of the system.

As shown in Figure~\ref{fig:architecturef}, the internal \texttt{Node.js} virtual machine is instrumented with a tracer that produces traces in Common Tracing File (CTF) format. For each environment, the native contexts and network communication functions are instrumented according to a formal event definition specification. The goal is to ensure traceability and deployment transparency for the developer.
In this specific case, microservices deployed with instrumented Java or \texttt{Node.js} container images do not require any developer intervention to configure the collection infrastructure. The effort associated with altering application source code through instrumentation is not needed. Likewise, as depicted in Figure~\ref{fig:architecturef}, an instrumented \texttt{Redis} image being part of the system is also included.

The proposed performance analysis framework aims to seamlessly integrate the various heterogeneous microservice components, enabling the developer to trace their system without changing the procedures in the application development and deployment pipeline.

The traces obtained in CTF format for each node are automatically sent to a trace aggregator. The synchronized traces obtained are loaded into Trace Compass (TC)\citep{compass} as an \emph{experiment} (a collection of related traces). TC extensibility allows for the development of analysis modules, enabling the creation of interactive tools that facilitate the understanding of the system operation.

In Figure~\ref{fig:architecturef}, under TC, the trace is read, and an event handler is used to extract various metrics and attributes from it. A trace provider constructs a powerful, efficient data structure to store event attributes. For each system state recorded by a particular trace event, its attributes and timestamp are used to build a data structure containing a history of the system states. Views, developed as extensions, allow querying the on-disk data structure through a data provider.
\subsection{Redis Cluster Collection Architecture}
\begin{figure*}[ht]
\centering{\includegraphics[width=10cm,height=7cm]{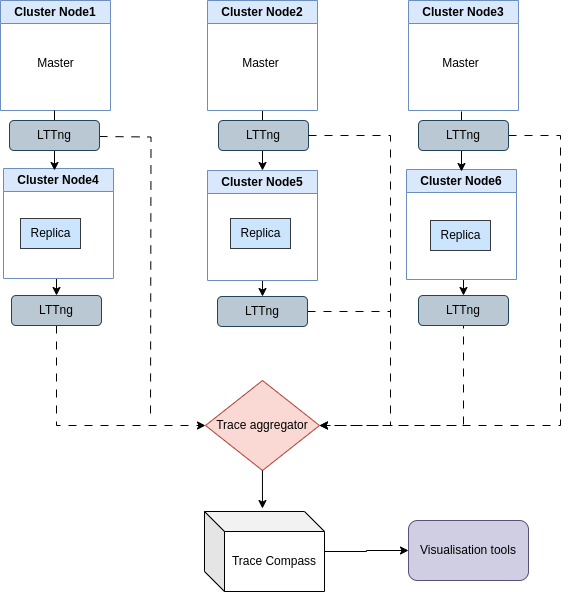}} 
\caption{Performance Analysis Integration Framework architecture}

\label{fig:architecture}
\end{figure*}

Figure~\ref{fig:architecture} illustrates the trace collection and analysis architecture. In this depiction of a \texttt{Redis}-based infrastructure, trace collection occurs through static instrumentation\citep{jia2013spire, serrano2009building}, during the initial phase of performance analysis. Then, the Linux Trace Toolkit next generation (LTTng) tracer, \citet{desnoyers2006lttng}, is activated in each of the nodes. These nodes can be containers, virtual machines, or physical hosts. Trace points are inserted at specific locations in the \texttt{Redis} code to collect the necessary events for analysis, aimed at debugging and diagnosing performance issues.
Within this collection infrastructure, the \texttt{LTTng} tracer generates events and stores them inside a CTF file on each node. Subsequently, these individual files are automatically aggregated by a trace aggregator, to obtain synchronized trace files. In the second step, which is the analysis phase, the traces are loaded as an experiment into Trace TC. In the latter, complex and automated analyses defined by algorithms are executed to identify specific events, correlate their metrics, and produce visualization tools focused on the ongoing performance analysis.
Conducting the analyses involves the creation of an advanced execution model, leveraging state-system technology and relying on attributes extracted from trace data. When reading the trace, the State History Tree (SHT), a highly efficient data structure, comes into play for the storage of the extracted attributes, along with additional analysis-related data. The TC framework supports the development of these data structures, benefiting from an optimization technique that enables rapid model queries in logarithmic time. TC boasts a range of capabilities for organizing traces, such as filtering and abstraction.
\subsection{Redis resources-driven modelling}
\label{subsec:model}

\begin{figure}[htb]
	\centering
	\includegraphics[width=0.7\linewidth]{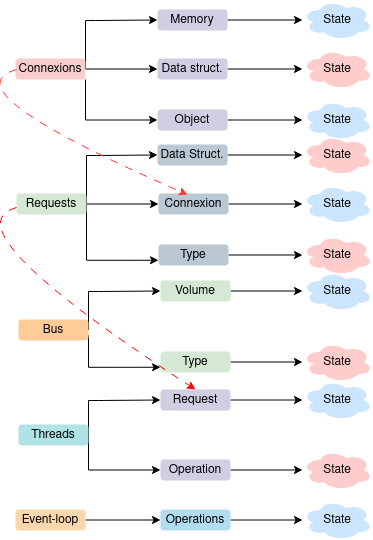}
	\caption{Model structure abstraction}
	\label{fig:model}
\end{figure}

The abstraction stage aims to develop a model that encapsulates the behavior of resources that could impact system performance. Specifically, five entities have been identified: \texttt{Connections, Requests, Bus, Threads, and Event-loop}. These entities serve as the entry points to the tree representing the model. The tree internal branches link to resources that define the states of different entities based on a timestamp. For instance, as depicted in Figure~\ref{fig:model}, the \texttt{Connections} entity is associated with three resources: \texttt{Memory, Data structure}, and \texttt{Objects}. These elements, linked to the \texttt{Connections} entity, abstract the resources potentially affecting system performance, within the context of network connections. For example, when a connection is received on a socket, \texttt{Redis} treats it as an incoming event, initially queued before being processed by the \texttt{Redis} event loop. 

The queue length is directly correlated with system performance \citep{gross2011fundamentals,geng2018exploiting,kasture2014ubik}. Memory-related attributes of the connection are attached to the \texttt{Memory} resource. A connection could be established by a client, be an internal link between two nodes in the cluster, or a link to a replica, and this information is attached to the \texttt{Type} resource state.

The \texttt{Requests} entity abstracts the resources potentially impacting performance in the context of requests execution. It includes three objects: \texttt{Data structure, type, and connection}. The \texttt{Data structure} object provides insights into the various data structures storing requests within the \texttt{Redis} infrastructure. A request might pass through several data structures, for example during message publication. The \texttt{type} object identifies the request type, which could be either \textit{read} or \textit{write}. The \texttt{Connection} resource links a \texttt{Connection} object to the request.

The \texttt{Bus} encapsulates the states of the network link between nodes in a \texttt{Redis} cluster. The \texttt{Volume} resource abstracts the volume of data transmitted over the Bus per second or millisecond. \texttt{Type} discriminates the nature of communication within the Bus, which could be \texttt{Primary} to \texttt{Replica} or \texttt{Node} to \texttt{Node}.

Regarding the \texttt{Threads} entity, it abstracts the various active threads involved in executing a particular operation. The \texttt{Request} resource identifies the request associated with the operation. The \texttt{Operation} resource attaches an ongoing operation executed by the thread. Therefore, information about the thread, the request, and the concerned operation can be obtained.

The abstraction of the event loop functioning is represented by the \texttt{Event-loop} entity, encompassing all operations necessary for monitoring its states.
The model construction involves creating a tree-like data structure replicating the model depicted in Figure~\ref{fig:model}. However, this phase occurs during the analysis of collected traces, which can contain a vast amount of events, millions or even billions. Efficient tools are necessary to support such analyses.

Trace Compass \citep{compass},  offers an ideal framework for building scalable, efficient data structures capable of handling large data volumes. This disk-based data structure, known as State History Tree (SHT) \citep{montplaisir2013state}, facilitates constructing the referenced model in Figure~\ref{fig:model}, and querying it for data visualization with logarithmic complexity. It also allows abstraction of the system state at a specific moment, and understanding the system various states over a time interval.

The model is populated using event attributes, and queries on the SHT return the system state at a precise timestamp. The states are the different values assigned to the model resource statuses, as presented in Figure~\ref{fig:model}. In other terms, the model can be viewed as an automaton, and its various states are preserved in the SHT with their timestamps.

\subsection{Trace analysis}
This section outlines the various analyses conducted on the trace to construct the model and generate the necessary visualizations for debugging performance issues in microservice architectures. Two types of analyses have been developed. The first analysis aims to efficiently traverse the trace to populate the model by creating a high-performance and scalable tree data structure. This process relies on different events from the trace and their attributes. The resulting model can be queried to understand Redis operations and to debug performance issues in a granular manner.

The second analysis focuses on integrating microservice components based on well-defined events and their attributes. It facilitates linking the analysis, obtained during the model construction, with other previously developed analyses. In the context of this paper, we integrate analysis \citep{mbikayi2023toward}, 
developed under Node.js. This analysis enables the transparent tracing of microservice architectures built with Node.js, without the need for application instrumentation. The integration is conducted transparently, allowing the connection of microservice executions with the Redis analysis, resulting in a comprehensive debugging approach.\\
\subsubsection{Analysis model construction}
\label{subsec:metrics}
This subsection describes how events are handled in a trace, to build a model as defined in Section~\ref{subsec:model}. 
\begin{table*}
\centering
  \caption{A subset of tracepoints monitored by our tool}
  \label{tab:tracepoints}
  \begin{tabular}{ll}
    \hline
    Tracepoints &Description\\
    \hline
    \texttt{start\_read\_client\_query} & Triggered when \texttt{Redis} reads an incoming connection from the queue.\\  \hline
    \texttt{end\_read\_client\_query}&Triggered when \texttt{Redis} when \texttt{Redis} exits connection reading\\  \hline
    \texttt{write\_to\_client\_start}&Triggered when \texttt{Redis} starts writing response to a connected client\\\hline
    \texttt{write\_to\_end\_start}&Triggered when \texttt{Redis} ends writing response to a connected client\\\hline
    \texttt{ssl\_read}&Triggered when \texttt{Redis} reads SSL connection\\\hline
    \texttt{free\_client}&Triggered when \texttt{Redis} frees client resources\\\hline
    \texttt{cluster\_read}&Triggered when \texttt{Redis} reads message from cluster bus\\\hline
    \texttt{cluster\_process\_packet}&Triggered when \texttt{Redis} has finished reading packet from the bus\\\hline
    \texttt{call\_command\_start}&Triggered when \texttt{Redis} executes a command\\\hline
    \texttt{call\_command\_end}&Triggered when \texttt{Redis} has finished executing command\\\hline   
    \texttt{add\_file\_event}&Triggered when \texttt{Redis} enqueues a new event\\\hline
    \texttt{delete\_file\_event}&Triggered when \texttt{Redis} remove an event after processing\\
    
  \hline
\end{tabular}
\end{table*}

\begin{figure*}[ht]
\centering{\includegraphics[width=14cm,height=4cm]{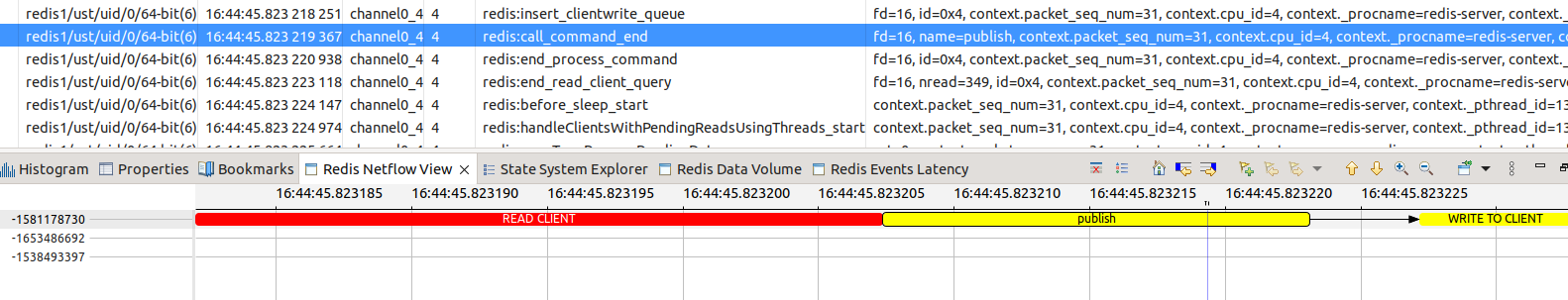}} 
    \caption{Partial view of the publish request life cycle. The top view shows the trace events. The bottom view shows the Netflow view used to track request life cycles. On the bottom left the TID are presented.}
    \label{fig:publish}
\end{figure*}

Consider a \texttt{Redis} client that transmits a request to publish particular data on a channel, to exemplify one facet on how the analysis operates. Figure~\ref{fig:publish} illustrates the outcomes of the analysis performed on the trace.

To run a command, the client establishes a connection with the server. The polling phase is executed on the \text{Redis} server side, during which the EL is notified of an incoming event on the socket. Following this, the \texttt{start\_read\_client\_query} tracepoint is triggered. 

During the trace analysis, the algorithm extracts the FD, the timestamp, and the thread ID immediately upon encountering this event. Subsequently, it stores these values in the SHT. In the model, \texttt{Connection} and \texttt{Requests} branches are instantiated. 

The FD is added to the \texttt{data structure} resource in the \texttt{Connexion} branch in the SHT, at that timestamp. Based on the FD, in the connection context, the \texttt{Memory} resource is populated with the value of the consumed memory, as determined by the event attributes.

The \texttt{Type} attribute of the corresponding resource is set to "client" in order to denote that the connection pertains to a client link. The FD is also added to the \texttt{Connection} resource within the \texttt{Requests} branch in the SHT, along with the timestamp extracted from the attributes.

Prior to encountering the \texttt{end\_read\_client\_query} event, indicating that the incoming event has been read from the connection, the algorithm watches the \texttt{call\_command\_start} event. The latter enables the decoding of the pending command and the invocation of the \texttt{Redis} function that is accountable for carrying out the execution. The timestamp and the command name are obtained by extracting the attributes from this event. 

The automaton then enters the \textit{"publish"}  state, as depicted in Figure~\ref{fig:publish}. As the analysis progresses, it encounters the serial \texttt{add\_file\_event} event, prior to reaching the \texttt{call\_command\_end} event. The \texttt{add\_file\_event} tracepoint instruments the queuing process of the event, for its execution in the subsequent phase of the EL.

The analysis then updates the \texttt{Data structure} resource state in the \texttt{Requests} branch, by adding the new request with it timestamp. Given the specific context in which the command to be executed is \textit{"publish}, the value \textit{"write"} is appended to the \texttt{Type} resource, to signify that the request pertains to a write operation. 

The \texttt{call\_command\_end} event, which is matched with the initial \texttt{call\_command\_start} event based on the FD, is used by the analysis to indicate the end of the command decoding and execution. The \texttt{end\_read\_client\_query} event, which is matched with the preceding \texttt{start\_read\_client\_query} based on the FD, marks the end of the connection reading.

Every alteration, made to the state of a resource in the model, is reflected as a state transition in the automaton. Throughout the process of the model construction, the system undergoes several transitions. A timestamp is assigned to each state, which matches a distinct event recorded in the trace. The SHT retains a history of the sequence of system states and transitions. Hence, it is possible to ascertain the state of the system at a specified timestamp by querying the SHT.

Upon reading from the connection, \texttt{Redis} acknowledges the client. The tracepoint \texttt{write\_to\_client\_start} is triggered whenever a message is transmitted from the server to the client. As the connection is not yet withdrawn from the context, the analysis modifies the system state to point out the beginning of the client response. The FD, obtained from the \texttt{write\_to\_client\_start} event, is used to identify the SHT branch position that corresponds to the initial client connection. 

Figure~\ref{fig:publish} depicts the system move to the \textit{"Write to client"} state. The algorithm uses the \texttt{write\_to\_client\_end} event to signify the end of the response transmission to the client. The FD is used to match the two events. Finally, the constructed data structure is used by the state system to provide a snapshot of the system state at a specific timestamp.

The same example can be used to show how the \texttt{Bus} branch is populated during the analysis. The bus defines the state of network communication within a \texttt{Redis} cluster, and is populated by the \texttt{cluster\_send} event, that is triggered when Redis broadcasts the publish message to the nodes on cluster. In this case, the status of the \texttt{Volume} resource in the SHT takes the value of the calculated volume of data sent metric. The \texttt{Type} resource is appended with the value of the communication type, in this case "broadcast". 

All the branches of the SHT are populated based on event attributes. The FD is a discriminator for different "Connections" and "Requests" objects. The \texttt{Data structure} resources store all the objects for a given branch. For instance, it stores all the connections represented by the FDs, in the case of the \texttt{Connections} branch.   

TC extensibility enables the development of extension modules that leverage the SHT as the data source for a variety of views. These extensions result in visual and interactive tools that facilitate the debugging of performance issues in \texttt{Redis}.\\
\subsubsection{Integration Analysis}

The integration process assembles various analyses to provide a comprehensive and detailed view of the system operation, facilitating performance analysis with a high level of granularity. For instance, this integration enables the visualization of interactions between different components of the microservice architecture as spans. However, it is crucial to be able to pinpoint the cause of a bottleneck when one is detected, whether due to a bug, a blocked process, or another issue. Delving deeper into a problem, at a finer granularity, is necessary to trace back to the root cause.

When a bottleneck is identified within Redis, the Redis analysis can be used to accurately determine the cause. Similarly, if the bottleneck is located within a Node.js microservice, the granular analysis of that environment can be leveraged for precise identification of the cause. The same applies to heterogeneous microservices developed with other technologies.

Integration offers a unified view of the system by observing interactions between microservices, while granular analyses of each environment enable tracing back to the root cause of a bottleneck.

In tracing components of the microservice architecture, events such as \texttt{http\_client\_request}, \texttt{http\_server\_receive}, \texttt{http\_server\_response}, and \texttt{http\_client\_response} are generated with each microservice interaction. These events capture the moments when a microservice sends a request to another, receives a request, sends back a response, and when the response is received by the sender, respectively.

The analysis relies on the attributes of these events to connect components. Attributes include source and destination addresses, source and destination port numbers, invoked services, and file descriptor numbers.

This information allows for the reconstruction and alignment of different interactions, connecting them with analyses developed for various environments, to seamlessly integrate them into a unified analysis tool.

While we used two environments (Node.js and Redis) in our case, this approach can be extended to various environments, such as Java, Golang, and others.

\section{Experiments}
\label{sec:exp}
This section introduces the context of our experiments to validate our approach. We present the environment configuration for the running the experiments. We then introduce practical and real use cases of performance issues encountered by the community. Our tool is subjected to these use cases to validate its effectiveness.

Then, we measure the overhead incurred by our approach. This measurement is crucial to determine its applicability.
\subsection{Environment configuration}
The experiments  were conducted with \texttt{Node.js} versions 12.22.3 and 16.1.0 with \texttt{Redis} 7.0.11, on an I7 Core running Ubuntu 20.4 with 16 GB RAM.
\subsection{Use cases}
\subsubsection{Multi-plateform Analysis Integration}
In this use case, to demonstrate the efficiency and advantages of our integrated performance analysis approach, the application \textit{MICROSERVICE-RESTful-Nodejs}\footnote{\url{\urlA}} was deployed and extended to add additional features, such as message publishing for subscribers.

The microservices application is developed in \texttt{Node.js} and uses \texttt{Redis} as a cache. It was modified to include message broker functionalities. It consists of five microservices organized as follows:
\begin{itemize}
    \item The "user" microservice manages user-related requests within the system.
    \item The "user" microservice manages user-related requests within the system.
    \item The "Order" microservice handles different orders made by users.
    \item The "Gateway" microservice receives user requests and redirects them to the appropriate microservice.
    \item The "Redis Gateway" microservice is contacted by other microservices looking to interact with Redis. It forwards the requests to the Redis server and returns responses to the relevant microservices.
    \item The "MySQL Gateway" microservice is responsible for managing various requests with the MySQL database.
\end{itemize}

\begin{figure*}[ht]
\centering{\includegraphics[width=11cm,height=6cm]{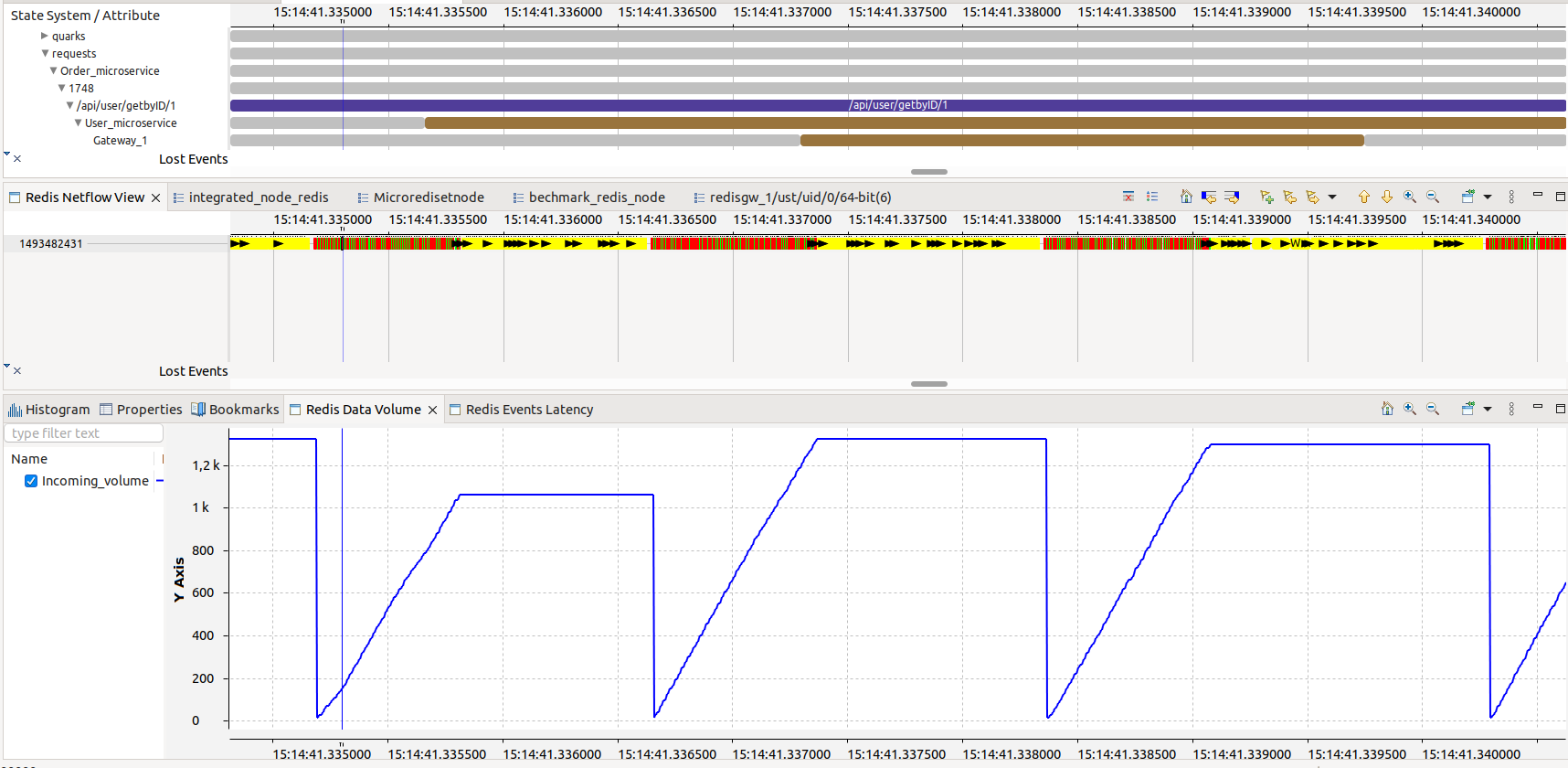}} 
    \caption{Constructed blocks from different analysis}
    \label{fig:combined}
\end{figure*}

To enable the assembly of blocks from the two different environments (Node.js and Redis) for performing granular performance analysis, the original \texttt{Redis} and \texttt{Node.js} images were replaced with instrumented images.

In the case of \texttt{Node.js}, \citet{mbikayi2023toward} proposed a low-overhead approach for transparent tracing for Node.js-based microservices. The reader is referred to their work for more details.

Running the application generates local traces in each node of the system. These traces are aggregated and loaded into TC. During the loading process, algorithms performing optimized analyses handle trace events, extract metrics from their attributes, and build a model on disk for each assembled environment.

The execution of the analyses produces two concurrent models under TC, one related to \texttt{Node.js} and the other to \texttt{Redis}. The extensions of the views generated by the various analyses allow them to be combined into a unified, integrated, and precise tool for performance analysis. The granularity is defined by each of the analyses developed for the different environments.

An overview of the outcomes of the performed analyses is presented in Figure~\ref{fig:combined}. In the latter figure, the top perspective pertains to operations executed in \texttt{Node.js}. It can be seen how the integration analysis aligns all executions. The lower granular perspectives pertain to operations executed in \texttt{Redis}. Notably, these outcomes are acquired in a transparent manner, devoid of any involvement from the developers. The system deployment, tracing, and analysis are all carried out in a seamless fashion. The responsibility of the developer is to accurately determine the nature of the performance issue by using the interactive tools that are generated.\\
\subsubsection{Leveraging Model Abstraction for Redis cluster request life-cycle tracking}
The present use case sufficiently demonstrates one of the major features of our tool, which allows tracking the life cycle of a request through a Redis infrastructure. Available monitoring and performance tracking tools primarily rely on established metrics to detect performance issues in Redis. These metrics may include disk write speed, the number of queries, memory-related information, to name a few. However, in a Redis infrastructure, such as a cluster, it may be essential to trace the execution path of a request to precisely identify where and when a request might be blocked, if necessary. This requires correlating information from all nodes in the cluster, synchronizing them using efficient analytical methods, to establish the path a request follows and visualize its journey through various nodes until the end of its life cycle.

This capability potentially allows for the precise localization of a performance issue occurring on the critical path of a request. To demonstrate this unique feature of our approach, we modified the Redis-bench client to enable it to send various types of commands, such as "publish" and "subscribe." The benchmark is then executed with 20,000 requests for each command (get, set, subscribe, publish).
\begin{figure*}[ht]
\centering{\includegraphics[width=15cm,height=2.5cm]{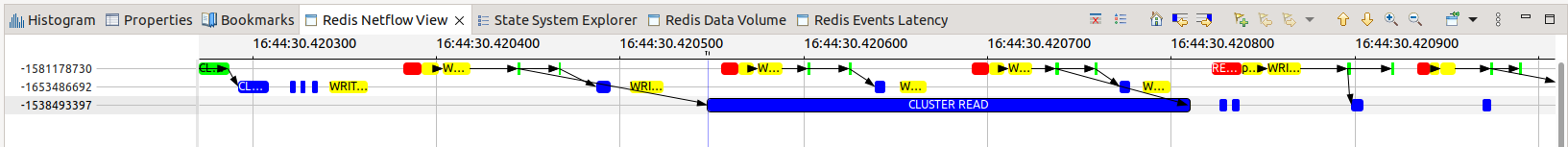}} 
    \caption{Redis cluster request flows}
    \label{fig:flow}
\end{figure*}
Figure~\ref{fig:flow} illustrates the flow of some requests processed in the Redis cluster. A performance issue can be observed, occurring during the reading of a message transmitted from one node to another. The read state is represented in the blue segment. When compared to various other requests, reading a message in the cluster takes less time.

\begin{figure*}[ht]
\centering{\includegraphics[width=15cm,height=4cm]{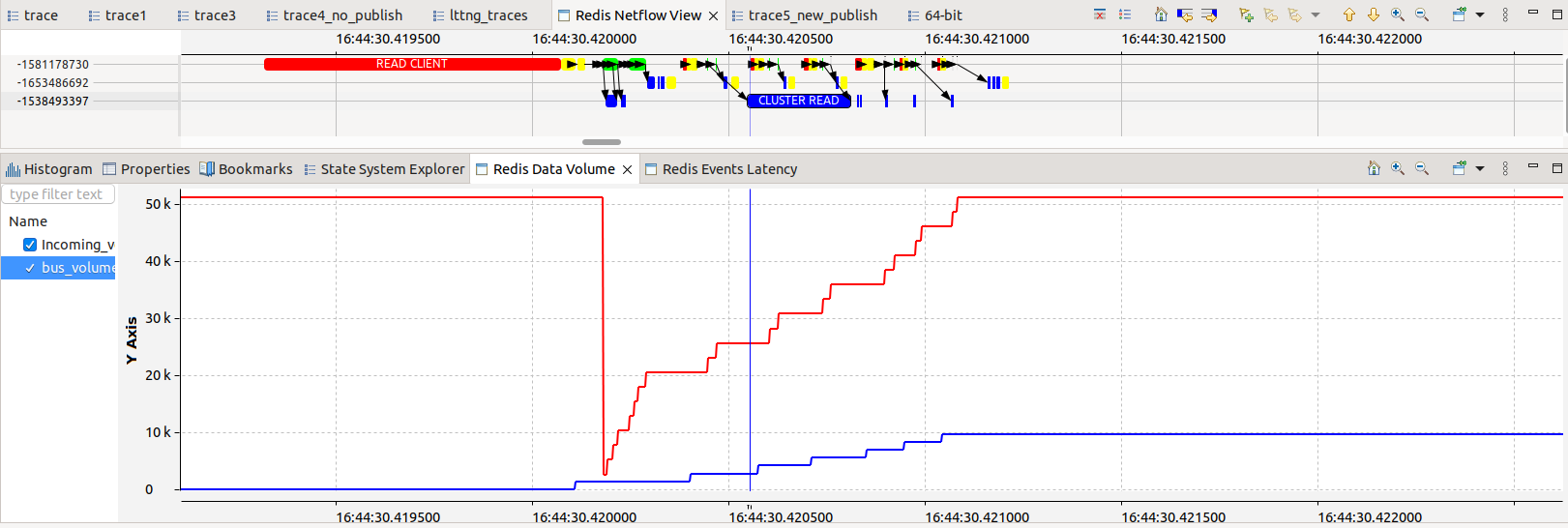}} 
    \caption{Request traveling on the cluster bus. Colors represent the operations performed on a timeline}
    \label{fig:readperf}
\end{figure*}

By parallelizing the Redis Netflow view, which shows the life cycle of requests, and the Redis Data Volume view, a correlation is observed between the performance issue, and a significant increase in the volume of data leaving through the bus. See Figure~\ref{fig:readperf}. This directly suggests a relationship between the performance issue and the state of the queue at the network interface.

Such a view also allows visualizing the actual execution time of a command, how long the request takes on each node visited, and how much time it takes within the cluster before reaching the client.\\
\subsubsection{Leveraging Model Abstraction in Redis Cluster Internal Communication}
In this use case, we demonstrate the effectiveness of our tool in detecting performance issues related to the communication within a Redis cluster. This specific case identifies a performance problem raised on GitHub by developers. This use case clearly demonstrates how our model level of abstraction, in terms of the communication bus, can be applied to conduct a performance analysis under Redis. The bug described on GitHub, which was still unresolved, pertains to the context of "publish" requests sent to a Redis infrastructure consisting of a cluster with a Primary/Replica or Primary/Primary configuration. Initial experiments indicate, for instance, that when 10KB of data is published in a Redis cluster, subscribers indeed receive 10KB of data, but the data exchanged between the cluster nodes is ten times greater. The use of a network monitoring tool confirms that the volume of data leaving the network interface imposes a very high overhead on the cluster.
\begin{figure*}[ht]
\centering{\includegraphics[width=15cm,height=4cm]{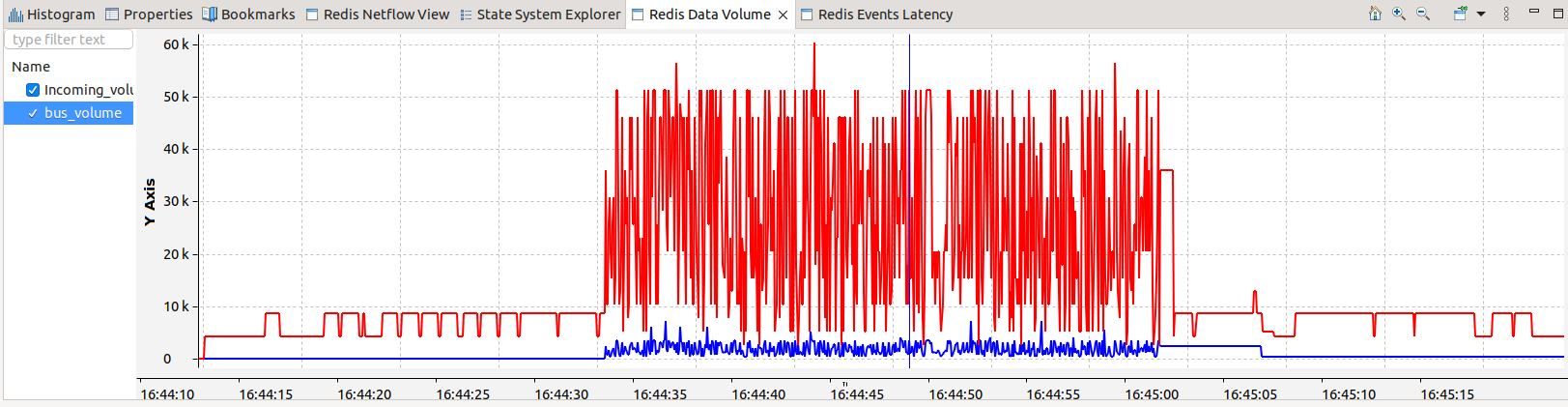}} 
    \caption{Redis cluster data volume traveling within. The bus is an abstraction of the communication link to multiple nodes}
    \label{fig:bus}
\end{figure*}
To reproduce the performance problem, a Redis cluster with $3$ nodes was configured. The microservice application "nextjs-express-redis-microservice-architecture" was used, along with \texttt{Jmeter} \citep{jmeter}, to generate users load for publishing messages to the cluster, with each client writing to a specific channel. The connection is established on Node $1$. On Node $2$, a client has subscribed to messages on the same channel, and is supposed to receive all the messages emitted by the publisher.

Figure~\ref{fig:bus} shows the result of the executed benchmark. The blue curve shows the progress of the volume of data entering the cluster over time, via the connection established by the client on Node $1$. This same volume of data is received by the client on Node 2. However, there is a drastic increase in the volume of data transmitted within the cluster bus, as indicated by the red curve. 
\begin{figure*}[ht]
\centering{\includegraphics[width=15cm,height=3cm]{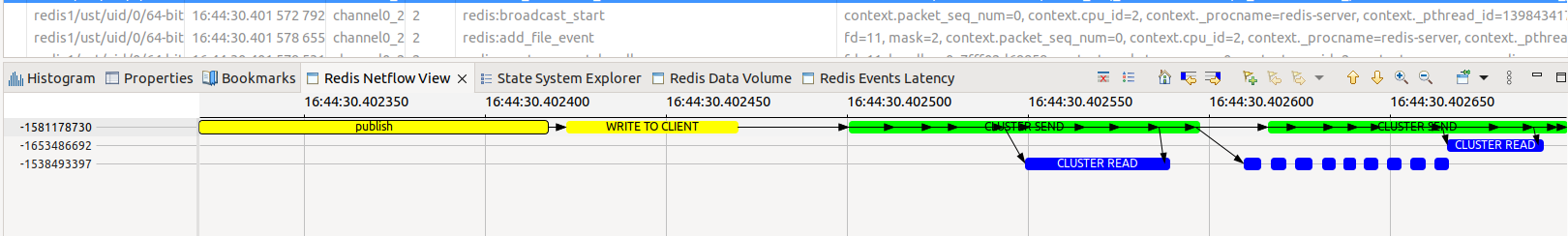}} 
    \caption{Redis Node is broadcasting the same message to the two other nodes}
    \label{fig:broadcast}
\end{figure*}
Further investigation reveals that the overhead imposed on the data volume increases linearly with the number of nodes connected to the cluster. Two elements significantly impact the volume of data exiting the interface. First, Redis uses a specific protocol called Gossip to encode communications between the cluster nodes. The encapsulation of the data received at Node $1$ is done using \texttt{Gossip}, which inherently introduces a very high overhead for small-sized data. The second element affecting the volume of data exiting the interface is that the communication on the bus is performed for each "publish" request in a broadcast manner. In other words, Node $1$ broadcasts the same message to all nodes connected to the cluster, as depicted in Figure~\ref{fig:broadcast}.

This use case reveals a Redis performance problem inherent to the Gossip protocol it employs. Resolving this issue requires a reevaluation of its communication mechanism within the cluster.\\
\subsubsection{Leveraging Model Abstraction to Expose Resource Conflicts}
This use case reveals a performance issue that occurs when sending a request containing a large item, in multiple parts, to \texttt{Redis}. The problem arises when \texttt{Redis} is compiled with \texttt{TLS} support and is run with multi-threading enabled for I/O operations. 

To reproduce the performance problem, traffic is subsequently generated using \texttt{Memtier}\citep{memtier}, which supports TLS, to execute multiple concurrent requests on \texttt{Redis}. This results in the activation of threads to handle various I/O operations. During the benchmark execution, sending an isolated request with a large volume of data arriving in parts forces \texttt{OpenSSL} to have remaining bytes in the buffer. This automatically creates a performance problem that leads to a \texttt{Redis} crash.
\begin{figure*}[ht]
\centering{\includegraphics[width=15cm,height=2cm]{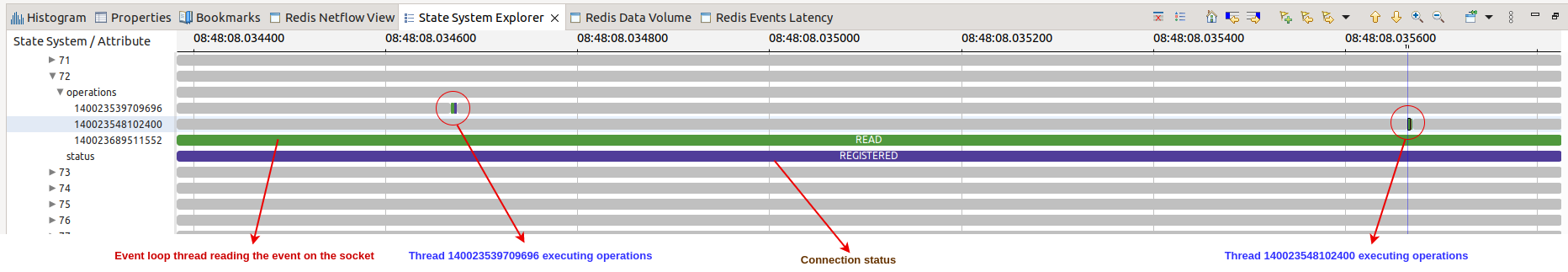}} 
    \caption{Persistent Model \texttt{Threads} branch view. Numbers on the left represent connections FDs.}
    \label{fig:freedesc}
\end{figure*}
\begin{figure*}[ht]
\centering{\includegraphics[width=15cm,height=2cm]{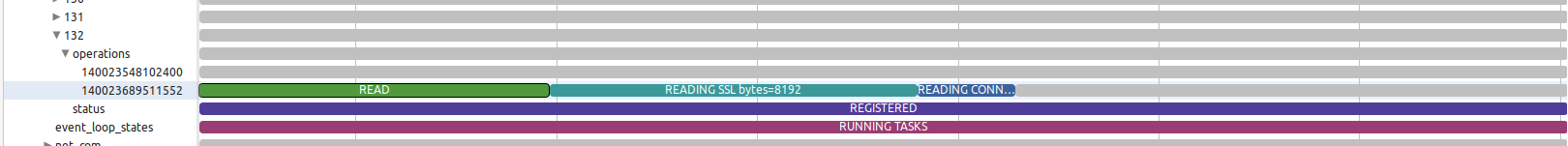}} 
    \caption{First reading loop of SSL connection}
    \label{fig:ssl1}
\end{figure*}
\begin{figure*}[ht]
\centering{\includegraphics[width=15cm,height=2cm]{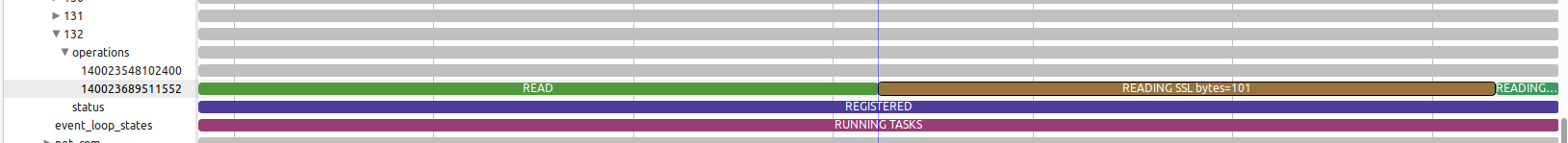}} 
    \caption{Second reading loop of SSL connection}
    \label{fig:ssl2}
\end{figure*}
Our approach accurately exposed the sequence of events that leads to this performance problem.
As can be observed in Figure~\ref{fig:freedesc}, which shows a portion of the Threads branch of the model, the view allows the visualization of the history of different established connections stored in the SHT. The numbers on the right represent file descriptors. In \texttt{Redis}, events arrive under different connections where they are read and processed. At the end of their processing, the events become unreadable only by being removed from the context of the connection to which they belonged. This operation is instrumented by the tracepoint \texttt{delete\_file\_event}. 

However, the connection remains open and can be reused for reading various other events, unless it is explicitly closed or due to abnormal behavior in \texttt{Redis} execution. By carefully observing Figure~\ref{fig:freedesc}, it can be seen that, for the connection identified by number $72$, the \texttt{Operations} attribute constitutes the logical resource associated, attached as per the model description. In other words, the resource is not physical but rather logical in nature in this instance. In this case, the various operations executed on this connection will be associated with the thread that performs them. 

A particular thread that draws attention is the one responsible for running the event loop. As described in Section~\ref{fig:model}, the EL reads incoming events on the connection, decodes the command to be executed, and queues it for processing in its next phase. The \textit{"Read"} state indicates the event reading by the thread, while the status described as \textit{"Registered"} indicates the state of the connection handler.

Based on the foundations laid out above, a description of the unique features of our tool, in accurately pinpointing the root cause of the bug, is given.
\begin{figure*}[ht]
\centering{\includegraphics[width=14cm,height=2cm]{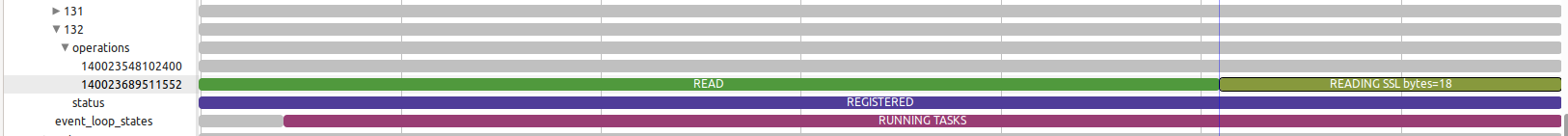}} 
    \caption{Third reading loop of SSL connection}
    \label{fig:ssl3}
\end{figure*}
\begin{figure*}[ht]
\centering{\includegraphics[width=14cm,height=2cm]{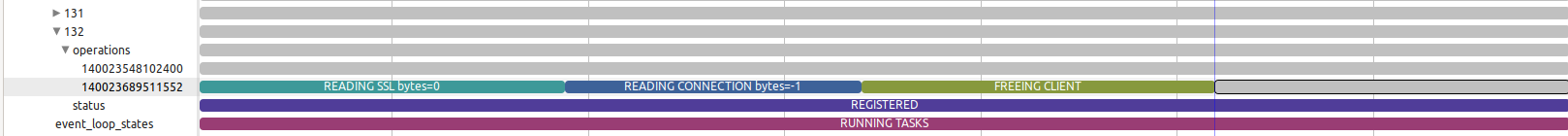}} 
    \caption{Fourth reading loop of SSL connection. No more data in the buffer, returns $-1$. Frees the connection.}
    \label{fig:ssl4}
\end{figure*}
\begin{figure*}[ht]
\centering{\includegraphics[width=14cm,height=2cm]{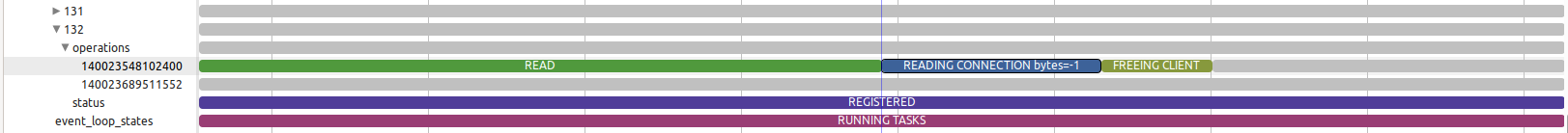}} 
    \caption{Fifth reading loop of SSL connection by another thread. No pending data in buffer, double-frees the connection}
    \label{fig:ssl5}
\end{figure*}
This particular use case illustrates a \texttt{Redis} bug concerning the \texttt{TLS} support with \texttt{OpenSSL}. When a big item is transmitted to \texttt{Redis} in numerous elements, \texttt{Redis} optimizes its operations by exclusively requesting the data required to finish the reading. Consequently, it prevents the creation of extra copies. On occasion, this optimization strategy may results in {OpenSSL} buffers retaining bytes.

\texttt{Redis} features mechanisms for periodically determining whether pending data is present in the buffer to mark the connection as such. When the multi-threading reading option is activated, threads from the pool will execute the connection reading. Given that the connection is identified as having pending data in the buffer, the next phase of the EL will entail a thread reading the connection to retrieve the pending data. Once no further data is available for reading, the connection will no longer be identified as such. It can thus be reused in response to subsequent incoming events.
\begin{figure*}[ht]
\centering{\includegraphics[width=14cm,height=2cm]{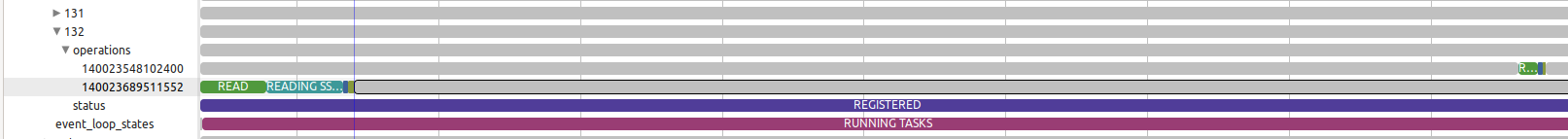}} 
    \caption{An zoom out of how the  operations performed by the threads.}
    \label{fig:zoom}
\end{figure*}
\begin{figure*}[ht]
\centering{\includegraphics[width=14cm,height=2cm]{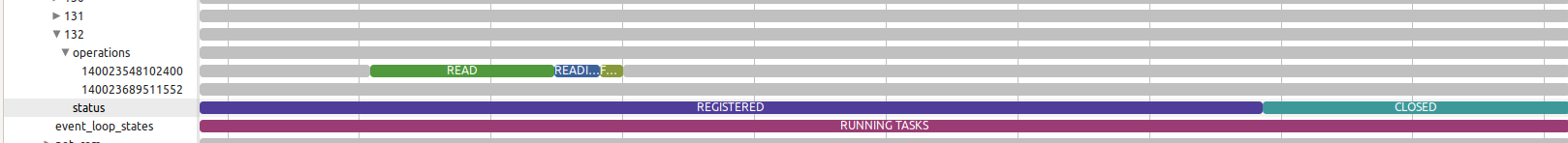}} 
    \caption{Fifth reading loop of SSL connection}
    \label{fig:close}
\end{figure*}

In the specific case of our experiment, the forced closure of the connection on FD $132$ triggers a new read event on the connection. Figure~\ref{fig:ssl1} depicts the reading of the client request by the EL. Since TLS support is enabled, the SSL handler calls the internal SSL reading function within \texttt{Redis}. A detailed look on Figure~\ref{fig:ssl1} captures the state of the thread transitioning to \textit{"Reading SSL bytes=8192"}. A straightforward deduction provides more precision regarding the amount of data that \texttt{Redis} requested from \texttt{OpenSSL} during the connection reading. In this case, $8192$ bytes are read. After the reading, the EL checks whether additional data is still pending in the buffer. If so, the connection is added to a list for processing in the \textit{"RUNNING TASK"} phase. In Figure~\ref{fig:ssl1}, the thread reading the connection is the main \texttt{Redis} process that manages the EL.

As soon as the EL enters the next \textit{"RUNNING TASK"} state, the tracepoint  \texttt{handleClientWithPendingReadsUsingThreads} is activated. This indicates the involvement of the I/O threads due to the activation of this support during \texttt{Redis} launch. Figure~\ref{fig:ssl2} depicts the second iteration of reading the connection. In this specific phase, the concerned thread traverses the list and generates read events for each of the connections with pending data in the buffer. The same connection FD $132$ is read again, as indicated by the thread status in Figure~\ref{fig:ssl2}. This time, the EL reads $101$ bytes in this second iteration. 

The check for pending data in the buffer is performed once more. In this case, the connection is again added to the list of connections with pending data for processing in the buffer. In the subsequent phase of the loop, represented by the \textit{"RUNNING TASK"} state, as shown in Figure~\ref{fig:ssl3}, the thread once again goes through the list and generates read events on each connection present. This time, $18$ bytes are read. In the fourth iteration, Figure~\ref{fig:ssl4}, since the quantity of data is not greater than zero, indicating that there is no more data pending in the buffer, the resources associated with the client represented by the connection are released. This can be observed in Figure~\ref{fig:ssl4} under the \textit{"FREEING CLIENT"} state. 

The connection is then removed from the list of connections with pending data in the buffer. However, it is not removed from the list of connections requiring the activation of threads for I/O processing. With the resources of this connection now released, the FD appears as readable. In the next phase of the EL (RUNNING TASK), given that there is no more pending data in the buffer for connection FD $132$, the activation of the I/O context, allowing for thread-based processing delegation, is not triggered, as there is no more data waiting to be read.

In the next cycle of the EL, since the connection is still readable, a read event is generated. This can be observed in Figure~\ref{fig:ssl5}. This time, a different thread comes into play and performs the reading on the connection, adding it to the list of connections requiring the activation of the thread context for I/O processing. With careful attention, this operation adds the same connection a second time to the thread I/O processing list. 

The consequence is that when there is no more data pending in the buffer, the thread frees the client connection for a second time, leading to the crash of \texttt{Redis}. A focused view of those operations is depicted by Figure~\ref{fig:zoom}. In response to this unexpected behavior, the connection is automatically closed, as depicted by Figure~\ref{fig:close}, and the sequence of events results in a system crash.

This problem exposes a synchronization flaw in \texttt{Redis} 7. Furthermore, we extended the experiment by introducing delays in the input/output thread reading operation, using a timer to simulate contention during the execution of the thread read callback. In some cases, this led to a race condition, detected by our tool, clearly showing that, in the same context, two different threads accessed the same connection, resulting in abnormal system behavior.
\subsection{Evaluation}
\label{sec:eval}
\subsubsection{Latency overhead}
\begin{figure*}%
    \centering
    \subfloat[\centering Latency in microseconds with tracing activated and with no tracing.]{{\includegraphics[width=6cm, height=3.7cm]{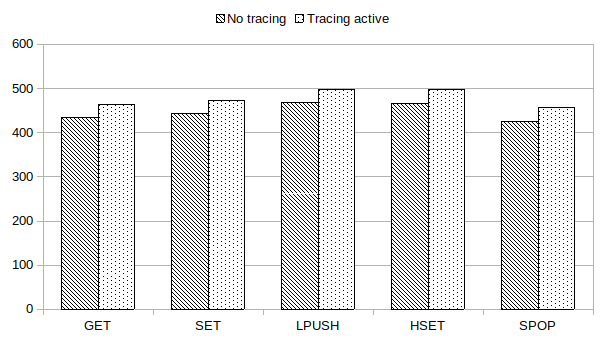} }}%
    \qquad
    \subfloat[\centering Resources utilisation with tracing activated]{{\includegraphics[width=6cm, height=3cm]{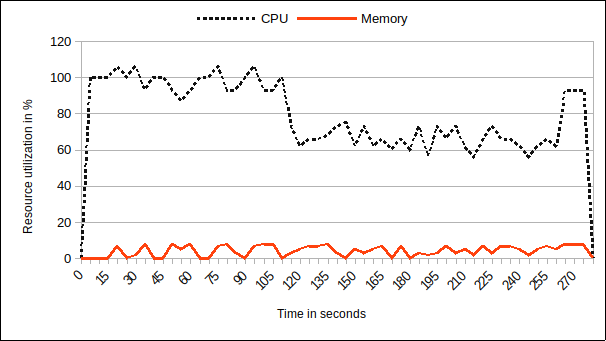} }}%
    \qquad
    \subfloat[\centering Resources utilisation with no tracing.]{{\includegraphics[width=6cm, height=3cm]{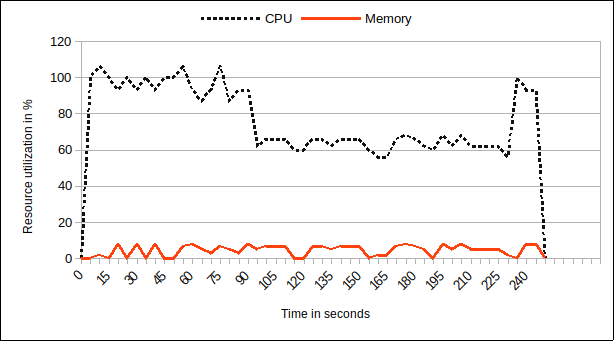} }}%
    \qquad
    \subfloat[\centering Analysis resources utilisation]{{\includegraphics[width=6cm, height=3cm]{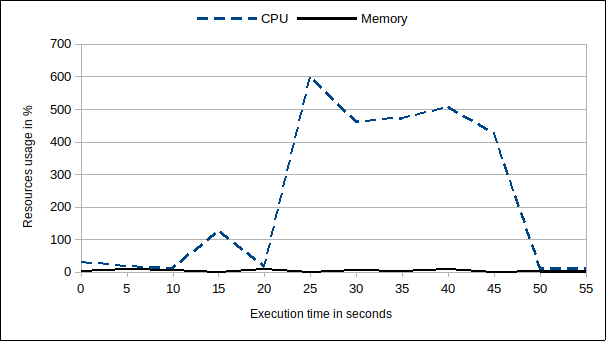} }}%
    \qquad
    \subfloat[\centering Analysis running time in second vs trace file sizes]{{\includegraphics[width=6cm, height=3cm]{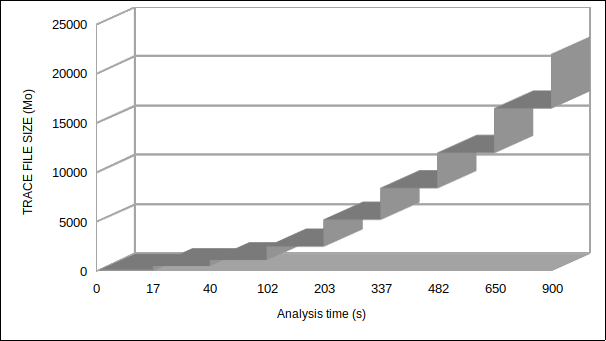} }}%
    \caption{Analysis resources utilisation}%
    \label{fig:overhead}%
\end{figure*}

The average execution time of various commands issued by the client is illustrated in Figure~\ref{fig:overhead}(a). Redis-benchmark was utilized to conduct the investigation. To assess the effectiveness of our solution in tracing, one million queries were sent to the server for each command on a local computer. A comparison between the execution of commands with and without tracing enabled reveals a disparity in the average execution time  (see Figure~\ref{fig:overhead}(a)). The overhead incurred by tracing is approximately  7\%. 

The fixed minimum guaranteed duration for processing the tracepoint by \texttt{LTTng} is the cause of this. However, the investigation is conducted directly on the internal processing of \texttt{Redis} and not in response to a request from a network service in a microservices context. \texttt{LTTng} is presently considered to be the world fastest tracer and imposes a minimal amount of system overhead.

\texttt{Redis} interacts with modules or external applications in the network, as part of its normal execution, to provide a highly available infrastructure. The fixed overhead introduced by tracing with our approach, which is approximately 70 microseconds in a configuration of concurrent command processing issued on 100 open connections, is absorbed by the communication time of microservice infrastructures.

\subsubsection{Resources Consumption}
The resource consumption of the executed  benchmark is illustrated in Figure~\ref{fig:overhead}(b) and Figure~\ref{fig:overhead}(c), corresponding to its usage with and without trace activation. It is clear that, on average, the curve does not impose a substantial burden on the system. This is due to the optimization of the instrumentation of the application, and the type of tracer employed. Comparatively to the non-traced application, the CPU consumption curve for the traced application exhibits a comparable pattern. The memory consumption curves exhibit a nearly identical profile in both scenarios, indicating that our solution has minimal effect on resource utilization.

\subsubsection{Trace Analysis}
Trace analysis is a critical component of our methodology, as it enables applying algorithms that have been designed to discover correlations among metrics derived from event attributes. It facilitates the creation of a data structure that is stored on disk and that contains the built model. This data structure is then used by the visualization tools needed for studying system performance. 

In Figure~\ref{fig:overhead}(d), the resources consumption incurred by running the analysis is depicted. To speed up the process, TC leverages a multi-threading approach, which explains the peaks in CPU consumption. However, memory is not much affected, since the designed data structure, storing the model data, is written to the disk. Figure~\ref{fig:overhead}(e) presents the time it takes to perform the analysis based on the trace file size. The analysis time is the duration necessary to parse a provided trace and construct the data structure that stores the model on disk. This data structure can then be queried by implemented views, in order to display correlated information, facilitating an interactive comprehension of the operation of the system.

\section{Conclusion}
The evolving landscape of applications, and the continuous growth of digital resources, have led to a paradigm shift in how content and information are accessed. Real-time interactions and minimal latency have become critical requirements for modern applications, leading to the adoption of cloud-based architectures and in-memory data storage solutions. \texttt{Redis}, as an open-source, in-memory data storage system, has emerged as a key player in this ecosystem, offering exceptional performance, versatility, and low-latency data access.

Debugging performance issues, in distributed systems, being a complex endeavor, specialized tools, that can correlate events and metrics to debug performance issues, are required. 

This paper introduced a novel methodology for the integrated debugging of \texttt{Redis}-based microservices performance issues. It leverages user-level tracing and a two-level abstraction analysis model, based on the identification of actors directly impacting \texttt{Redis} performance. The methodology employs two advanced tools for achieving its results. It leverages the state system technology and an SHT to store historical states of the system. Thus, stored metrics and information could be correlated, to provide a comprehensive solution for performance debugging. 

Our presented technique offers a valuable resource for diagnosing and addressing performance issues in \texttt{Redis}, ensuring that applications can maintain optimal performance and deliver an excellent user experience in the current demanding digital landscape.

An interesting avenue for future work could be to extend the model, to consider different types of resources and actors for performance debugging of other system architectures.

\label{sec:conclusion}
 \bibliography{acmart}
\vskip -2\baselineskip plus -1fil
\begin{IEEEbiography}[{\includegraphics[width=1in,height=1.25in,clip,keepaspectratio]{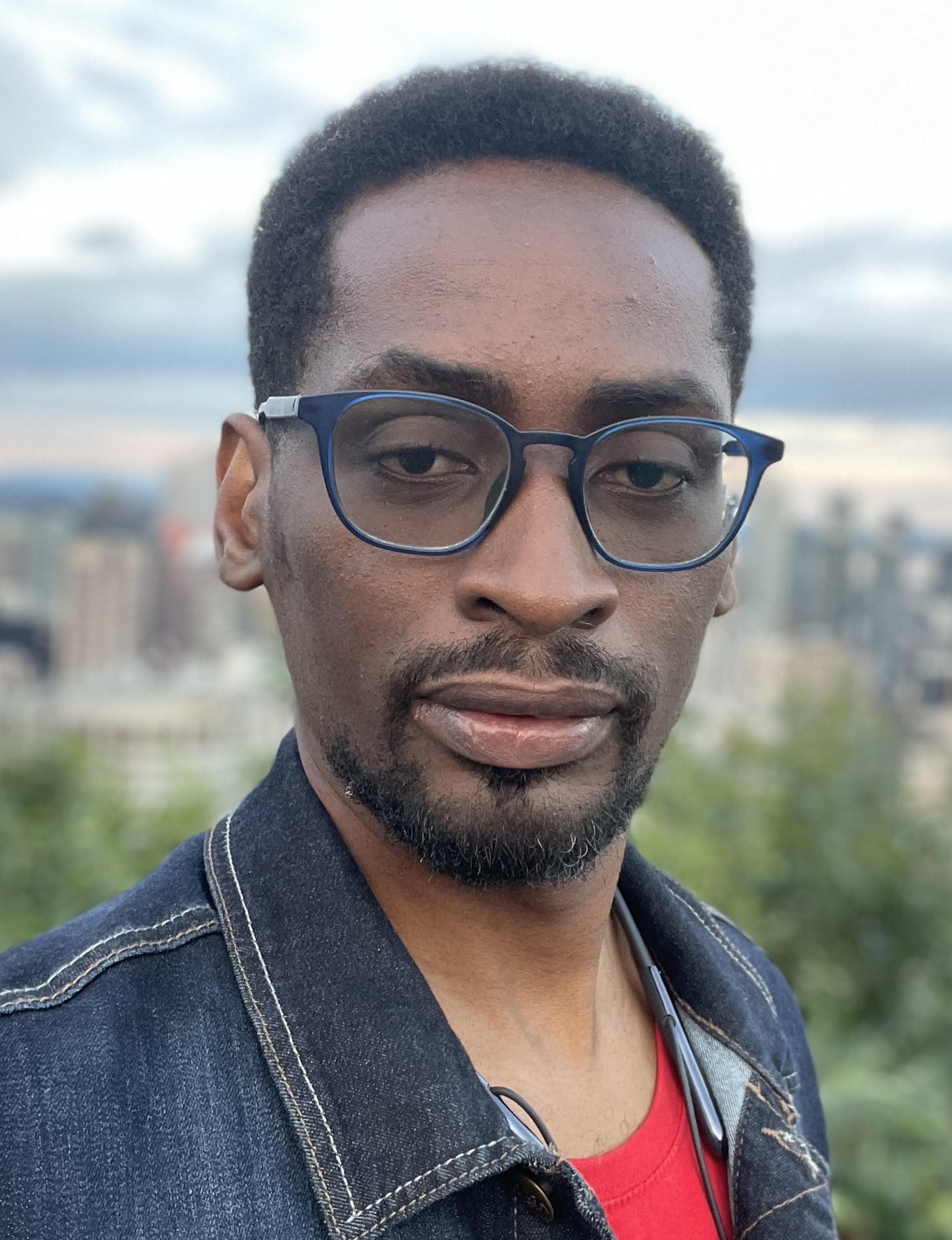}}]{Herve M. Kabamba}
is a PhD candidate at the Computer and Software Engineering department at Polytechnique Montréal, Canada. His research focuses on performance analysis of distributed systems using  tracing and monitoring approaches. He authored some publications related to the application of evolutionary algorithms for the resolution of complex problems in the field of computer security and signal processing. He obtained his Master's degree in 2020 from the Institut Supérieur de Formation Management, Ingenierie et Technologie of Dakar, Senegal in Engineering of Telecommunications Systems and Computer Networks. He received his bachelor degree in Computer Engineering in 2008 from the University of Kinshasa.  From 2006 to 2011 he held the position of Head of the IT department at Moneytrans RDC. His academic career began in 2011 where he held the position of university Assistant at the Institut Supérieur de Commerce de Kinshasa, then Senior Lecturer (Chef de Travaux) in 2015. He has since been a Lecturer and Academic Secretary of the computer science department.
\end{IEEEbiography}
\vskip -2\baselineskip plus -1fil
\begin{IEEEbiography}[{\includegraphics[width=1in,height=1.25in,clip,keepaspectratio]{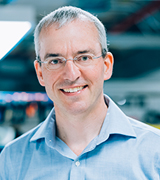}}]{Michel R. Dagenais} is professor at Polytechnique Montreal in the department of Computer and Software Engineering. He authored or co-authored over one hundred scientific publications, as well as numerous free documents and free software packages in the fields of operating systems, distributed systems and multicore systems, in particular in the area of tracing and monitoring Linux systems for performance analysis. In 1995-1996, during a leave of absence, he was director of software development at Positron Industries. In 1997, he co-founded the Linux-Quebec user group. Most of his research projects are in collaboration with industry and generate free software tools among the outcomes. The Linux Trace Toolkit next generation, developed under his supervision, is now used throughout the world and is part of several specialised and general purpose Linux distributions.
\end{IEEEbiography}
\vskip -2\baselineskip plus -1fil
\begin{IEEEbiography}[{\includegraphics[width=1in,height=1.25in,clip,keepaspectratio]{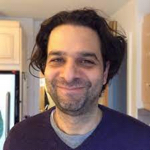}}]{Matthew Khouzam}
is the product manager for Eclipse Theia, OpenVSX, Eclipse Trace Compass within Ericsson and co-lead for the Eclipse Trace Compass Incubator. He is also co-lead of the CDT.cloud project. He is technology enthusiast, especially in the field of performance engineering. Matthew worked in academia and industry, coordinating between large and small companies as well as university research labs. He obtained his master's degree in computer engineering at Polytechnique Montreal in 2006, and his Bachelor degree in computer engineering at Concordia University in 2003.

\end{IEEEbiography}

\end{document}